\documentclass[3p,onecolumn,twoside,times]{elsarticle}

\usepackage{ecrc}

\volume{00}

\firstpage{1}

\journalname{Compte Rendus Mecanique}

\runauth{D.S.\ Goldobin}

\usepackage{graphicx}
\usepackage{amsmath}
\usepackage{amssymb}
\usepackage{mathtext}

\begin{document}

\begin{frontmatter}

\title{Non-Fickian Diffusion Affects the Relation between the Salinity
       and Hydrate Capacity Profiles in Marine Sediments}

\author{Denis S.\ Goldobin} 
\address{Institute of Continuous Media Mechanics, UB RAS,
         1 Acad.\ Korolev street, Perm 614013, Russia}
\address{Department of Mathematics, University of Leicester,
         University Road, Leicester LE1 7RH, UK}
\ead{Denis.Goldobin@gmail.com}

\begin{abstract}
On-site measurements of water salinity (which can be directly
evaluated from the electrical conductivity) in deep-sea sediments
is technically the primary source of indirect information on the
capacity of the marine deposits of methane hydrates.  We show the
relation between the salinity (chlorinity) profile and the hydrate
volume in pores to be significantly affected by non-Fickian
contributions to the diffusion flux---the thermal diffusion and
the gravitational segregation---which have been previously ignored
in the literature on the subject and the analysis of surveys data.
We provide amended relations and utilize them for an analysis of
field measurements for a real hydrate deposit.
\end{abstract}



\end{frontmatter}

\section{INTRODUCTION}
Since being discovered methane hydrates have attracted significant
attention. First estimates of their amount on the Earth,
especially in the marine sediments, and their importance were
extremely exaggerated. Today's assessment of their amount and role
are more moderate and well underpinned by field data and results
of numerical modelling and, thus, may be treated as realistic.
Even with this ``moderate'' evaluation, research on natural
hydrates in marine sediments is considered to be important.

In particular, methane hydrates present a potential hazard under
anthropogenic climate change. The sensitivity of hydrate stability
to changes in local pressure-temperature conditions and their
existence beneath relatively shallow marine environments, mean
that submarine hydrates are vulnerable to changes in bottom water
conditions (e.g.\ warming). The potential climate impact of
methane release following dissociation of hydrate in the past has
been compared to climate feedbacks associated with the terrestrial
biosphere and identified as a possible trigger of abrupt climate
change (e.g.,~\cite{Maslin-etal-2010,geohaz}). The role of hydrate
disassociation as a trigger for submarine landslides has also been
investigated~\cite{Maslin-etal-2010,hydr-slides-1,hydr-slides-2},
with reports of known hydrate occurrences that coincide with
slumping and submarine landslides being
common~\cite{hydr-slides-2}. It is therefore imperative to improve
our understanding of the global hydrate inventory.
Studies~\cite{Goldobin-Brilliantov-2011,Goldobin-2011} highlight
that this improvement requires also certain revision of the
physical and mathematical models of the marine deposits of methane
hydrates employed in the
literature~\cite{Davie-Buffett-2001,Davie-Buffett-2003a,Davie-Buffett-2003b,paradigm_models}.

\begin{figure}[!t]
\center{
\includegraphics[width=0.45\textwidth]%
 {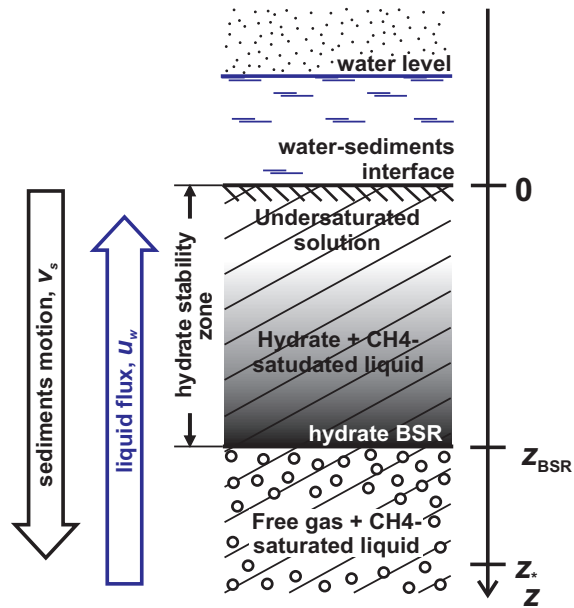}
}

  \caption{(Color online)
Sketch of the marine sediments hosting a hydrate deposit with a
free gas zone ({\it or} ``bubble horizon'') beneath the zone of
the thermodynamic stability of methane hydrate. The boundary
between the hydrate deposit and the free gas zone forms the bottom
simulating reflector (BSR) of acoustic waves and can be
seismically detected.}
  \label{fig1}
\end{figure}

Acquiring samples with methane hydrates from sediments beneath
deep water bodies is a costly procedure which is not practically
employed for large scale surveys~\cite{ODP}. Instead, the presence
of hydrate is typically inferred from seismic data
(e.g.,~\cite{Ecker-Dvorkin-Nur-1998}) or on-site salinity
measurements in boreholes (e.g.,~\cite{Davie-Buffett-2001}). The
seismic data are (i) the presence of the ``bottom-simulating
reflector'' which appears when hydrate deposit touches the bottom
boundary of the hydrate stability zone and, therefore, is
underlaid by a free gas
horizon~\cite{ODP,Davie-Buffett-2001,paradigm_models}
(Fig.\,\ref{fig1}), and (ii) sound speed increase in sediments
with hydrates owing to the sediments cementation by hydrate in
pores~\cite{Ecker-Dvorkin-Nur-1998}. Both seismic techniques have
significant limitations. For instance, the bottom simulating
reflector appears only when hydrate deposit reaches the bottom
edge of the hydrate stability zone, and the velocity increase
owing to the sediment cementation does not allow accurate
estimation of the amount of hydrate. Hence, on-site salinity
measurements become an important source of information.

To date, the mathematical models reconstructing hydrate deposit
parameters by means of fitting the measured salinity profiles
disregard non-Fickian contributions to the diffusion flux of salt
in sediments. In the present paper we (i) derive relations between
the profiles of the hydrate volumetric fraction in pores and the
measured salinity and (ii) demonstrate the non-Fickian
contributions to be important.

The paper is organized as follows. In Sec.\,2 we describe the
transport processes in carbon-rich marine sediments and derive the
relation between the salinity (chlorinity) and hydrate profiles.
In Sec.\,3 this relation is employed for reconstruction of the
hydrate profile for a real natural hydrate deposit in the Blake
Ridge hydrate province. In the concluding section we discuss
importance of the non-Fickian diffusion and implementation of our
reconstruction procedure.

\section{MARINE SEDIMENTS HOSTING HYDRATE DEPOSITS:
         TRANSPORT PROCESSES}
Real geological systems are much more uniform along two directions
(say horizontal) than along the third direction (say vertical).
Hence, we consider a one-dimensional problem with vertical spatial
coordinate $z$ (Fig.\,\ref{fig1}). On the field scale, such
systems are featured by the temperature growth with depth
\begin{equation}
T=T_\mathrm{sf}+Gz\,,
\label{eq201}
\end{equation}
where $T_\mathrm{sf}$ is the temperature of the water-sediment
interface ({\it or} seafloor) and $G$ is the geothermal gradient.

Methane is generated from the sediments by anaerobic bacteria. If
the temperature is low enough and the pressure is high enough,
methane forms hydrate. However, the critical pressure for the
thermodynamic stability of hydrate depends on temperature nearly
exponentially, and the hydrostatic pressure, which grows linearly
with depth, cannot compensate the linear growth of temperature
$T=T_\mathrm{sf}+Gz$ below a certain depth $z_\mathrm{BSR}$. This
depth $z_\mathrm{BSR}$ is the bottom boundary of the hydrate
deposit: below this depth, hydrate is dissociated into water and
methane-gas bubbles (Fig.\,\ref{fig1}).

The major part of natural hydrates of hydrocarbons on the Earth is
a structure I clathrate of methane ($>99\%$). The elementary cell
of an ``ideal'' structure I clathrate is formed by 8 molecules of
$\mathrm{CH_4}$ and 46 molecules of $\mathrm{H_20}$, i.e., the
mass fraction of water in the clathrate
$K_\mathrm{H_2O}=23\cdot{18}/(23\cdot{18}+4\cdot{16})\approx0.866$.
For real hydrates the saturation of the structure with methane
molecules is slightly less than $100\%$---some clathrate cages are
not occupied by the methane molecule. In geological systems, the
occupancy does not necessarily correspond to the thermodynamic
equilibrium, because the hydrate was initially formed under the
thermodynamic conditions of the place from where the hydrate was
transported due to diverse geological processes. The relaxation
rate of the occupancy is determined by the molecular diffusion of
the methane molecules in hydrate clathrate (a solid matter) and is
commensurable with rates of geological transport processes
($100-1000$ kyears). On the one hand, the actual local occupancy
cannot be exactly evaluated from thermodynamical principles and
the equilibrium condition, while, on the other hand, it is always
higher than $95\%$~\cite{Circone-Kirby-Stern-2005}, i.e.\ very
close to the ideal value, $100\%$. Hence, we assume the ``ideal''
structure of hydrate, $K_\mathrm{H_2O}\approx0.866$.

Hydrate forming in pores consumes water from the brine while salts
remains in the brine. Therefore, salt concentration increases and
diffusion drives redistribution of the salt mass. This process
determines the formation of the salinity (chlorinity) profile.

Additionally, one should distinguish the on-site chlorinity,
$\omega_s$, and the measured chlorinity, $\tilde{\omega}_s$,
because the drilling procedure results in dissociation of hydrate
and the release of the hydrate water into the brine at the
measurement site~\cite{ODP}. Given the volumetric fraction of
hydrate in pores is $h(z)$, the mass of $\mathrm{NaCl}$ in the
unit volume of pores before the dissociation of hydrate,
$\omega_s\rho_w(1-h)$, equals the mass after dissociation,
$\tilde{\omega}_s(\rho_w(1-h)+K_\mathrm{H_2O}\rho_hh)$, where
$\rho_w=1000\,\mathrm{kg/m^3}$ is the water density,
$\rho_h=930\,\mathrm{kg/m^3}$ is the hydrate density. Hence,
\begin{equation}
\tilde{\omega}_s=\omega_s\left(1-\frac{K_\mathrm{H_2O}\rho_h}{\rho_w}h
 +\mathcal{O}\left(\left(1-\frac{K_\mathrm{H_2O}\rho_h}{\rho_w}\right)^2h^2\right)\right)
 \approx\omega_s\left(1-\frac{K_\mathrm{H_2O}\rho_h}{\rho_w}h\right)
 \equiv\omega_s(1-kh)\,,
\label{eq202}
\end{equation}
where
$$
k\equiv\frac{K_\mathrm{H_2O}\rho_h}{\rho_w}\approx0.805\,,
$$
the correction
 $\mathcal{O}[(1-k)^2h^2]=\mathcal{O}[(0.03\cdot h)\cdot h]$
for real systems, where $h$ rarely exceeds
$7\%$~\cite{Davie-Buffett-2001,paradigm_models}, is less than
$0.002\cdot h$ and can be neglected.

Since the salt is transported with pore water, we have to describe
the water mass transport in the system. The water mass is
transported as a part of hydrate, with sediments, and with the
brine, by the pore water flux. The downward transport of sediments
is significantly affected by the sediments compaction with
depth~\cite{Davie-Buffett-2001,ODP}; the porosity $\phi$
significantly decreases with depth according to the empiric law
\begin{equation}
\phi(z)=\phi_0\exp(-z/L)\,,
\label{eq203}
\end{equation}
where $L$ is the depth of $e$-folding of porosity. The mass
conservation law for sediments yields
\begin{equation}
v_s(z)=\frac{1-\phi_0}{1-\phi(z)}v_{s0},
\label{eq204}
\end{equation}
where $v_s(z)$ is the sediment motion velocity. This relation is
nearly not affected by the conversion of part of sediments into
methane. Indeed, the defect of the solid matrix volume owing to
methane generation is approximately $\Delta V_s=\Delta
m_\mathrm{CH_4}/\rho_s$ ($\rho_s\approx 2650\,\mathrm{kg/m^3}$ is
the sediment material density~\cite{Davie-Buffett-2001}), whereas
the production of hydrate $\Delta V_h$ from this mass of methane
is $\Delta m_\mathrm{CH_4}/K_\mathrm{H_2O}\rho_h$. The ratio
$\Delta V_s/\Delta V_h=K_\mathrm{H_2O}\rho_h/\rho_s\approx 0.04$
is small and thus $\Delta V_s$ related to methane generation is
negligible.

For water, the mass conservation law reads
\begin{equation}
\frac{\partial}{\partial t}\big(\phi(1-h-b)\rho_w+\phi h K_\mathrm{H_2O}\rho_h\big)
=-\frac{\partial}{\partial z}\big(\rho_wu_w\big)
 -\frac{\partial}{\partial z}\big(\phi h K_\mathrm{H_2O}\rho_h v_s\big)\,,
\label{eq205}
\end{equation}
where $b$ is the volumetric fraction of bubbles in the pore volume
($h=0$ beyond the hydrate stability zone, $b=0$ within it), $u_w$
is the brine filtration velocity. Following Davie and
Buffett~\cite{Davie-Buffett-2001,Davie-Buffett-2003a}, we consider
a steady-state situation and set time derivatives to zero. Hence,
$$
\rho_wu_w(z)+\phi(z)h(z)K_\mathrm{H_2O}\rho_h v_s(z)=\rho_wu_{w}(0)
$$
(hydrate is not present close to the water-sediment interface,
$h(0)=0$), and, substituting $v_s(z)$ from Eq.\,(\ref{eq204}), we
find
\begin{equation}
u_w=u_{w0}-k\phi h\frac{1-\phi_0}{1-\phi}v_{s0}\,.
\label{eq206}
\end{equation}

Given the brine filtration velocity $u_w$ is known
(Eq.\,(\ref{eq206})), one can evaluate the salt transport from the
mass conservation law;
\begin{equation}
\frac{\partial}{\partial t}\big(\phi(1-h-b)\rho_w\omega_s\big)
=-\frac{\partial}{\partial z}\big(\rho_w\omega_su_w\big)
 -\frac{\partial}{\partial z}J_{s,\mathrm{diff}}\,.
\label{eq207}
\end{equation}
The diffusive flux $J_{s,\mathrm{diff}}$ of the $\mathrm{NaCl}$
mass is contributed by the Fickian molecular diffusion flux and
non-Fickian diffusion fluxes---the thermal diffusion and
gravitational segregation (importance of which for the gas
transport in geological systems under consideration was previously
demonstrated~\cite{Goldobin-Brilliantov-2011}). The diffusive flux
reads~\cite{Bird-Stewart-Lightfoot-2007,Goldobin-Brilliantov-2011}
\begin{equation}
J_{s,\mathrm{diff}}=-\chi\phi(1-h-b)D_s\rho_w\omega_s
 \left(\frac{\partial}{\partial z}\ln{\omega_s}
 +\alpha_s\frac{\partial}{\partial z}\ln{T}
 -\frac{\tilde{\mu}_sg}{RT}\right)\,.
\label{eq208}
\end{equation}
The following notations are introduced here:\\
$\bullet$~$\chi$ is the tortuosity factor, which characterize the
effect of the pore morphology on the effective diffusivity of
species in pore fluid. For our system
$\chi=0.75$~\cite{Butt-etal-2000}.\\
$\bullet$~$D_s$ is the molecular diffusion coefficient in bulk
brine.\\
$\bullet$~$\alpha_s$ is the thermodiffusion constant.\\
$\bullet$~$g=9.8\,\mathrm{m/s^2}$ is the gravity.\\
$\bullet$~The universal gas constant
$R=8.314\,\mathrm{J/(mol\,K)}$.\\
$\bullet$~ $\tilde{\mu}_s=\mu_\mathrm{NaCl}-N\mu_\mathrm{H_2O}$ is
the effective molar mass of the pair of ions $\mathrm{Na}^+$ and
$\mathrm{Cl}^-$ in the aqueous solution, $N$ is the number of
water molecules in the volume occupied by this pair in the
solvent, which can be evaluated from the dependence of the
solution density on its concentration (see, e.g., Appendix A
in~\cite{Goldobin-Brilliantov-2011}):
$$
\tilde{\mu}_s=\frac{\mu_\mathrm{NaCl}}{\rho_w}
 \left.\frac{\partial\rho_\mathrm{solution}}{\partial\omega_s}\right|_{\omega_s=0}
 \approx 42\,\mathrm{g/mol}\,.
$$

In~\cite{Caldwell-1973} the thermodiffusion constant $\alpha_s$
and the molecular diffusion coefficient $D_s$ were measured for a
seminormal aqueous solution of $\mathrm{NaCl}$.  For the
thermodiffusion constant a sign inversion was observed near
$T_\mathrm{i}=12^\circ C$.  In the temperature range typical for
our system, $T\in(275\,\mathrm{K},305\,\mathrm{K})$, the
temperature dependencies of $\alpha_s$ and $D_s$ are strong and
well represented by expressions
$$
\alpha_s\approx 0.0246\,\mathrm{K^{-1}}(T-T_\mathrm{i})
\qquad\mbox{ and }\qquad
 D_s\approx 6.1\cdot 10^{-10}
 \exp[0.0371\,\mathrm{K^{-1}}(T-273.15\,\mathrm{K})]\,\mathrm{m^2/s}\,.
$$
Hydrostatic pressure in marine sediments, which is up to several
hundreds atmospheres, is not strong enough to affect the diffusion
constant of chemicals in water.

For a steady state, Eqs.\,(\ref{eq207}) and (\ref{eq208}) yield
\begin{equation}
 \omega_su_w
 -\chi\phi(1-h-b)D_s\omega_s
 \left(\frac{\partial}{\partial z}\ln{\omega_s}
 +\alpha_s\frac{\partial}{\partial z}\ln{T}
 -\frac{\tilde{\mu}_sg}{RT}\right)
=
 \omega_{s*}u_{w*}+J_{s,\mathrm{diff}*}\,,
\label{eq209}
\end{equation}
where $\omega_{s*}\equiv\omega_s(z_*)$,
 $u_{w*}\equiv u_w(z_*)=u_{w0}$,
 $J_{s,\mathrm{diff}*}\equiv J_{s,\mathrm{diff}}(z_*)$,
and $z_*$ is a certain depth deep below the hydrate stability zone
(see Fig.\,\ref{fig1}). Hereafter, the sign ``$*$'' indicates the
value at depth $z_*$.

Since $h\ll1$, we restrict our consideration to the linear in $h$
(and $b$) approximation. Substituting Eqs.\,(\ref{eq202}) and
(\ref{eq206}) into Eq.\,(\ref{eq209}) we can find
\begin{equation}
 \frac{\partial h}{\partial z}
 +\gamma(z)\,h
 =f(z)\,,
\label{eq210}
\end{equation}
where
$$
\gamma(z)=
 \frac{1}{\chi D_s}\left(\frac{1-\phi_0}{1-\phi}v_{s0}
 -\frac{u_{w0}}{\phi}\right)\,,
\quad
f(z)=
 -\frac{1}{k}\left(\frac{1}{\tilde{\omega}_s}\frac{\partial\tilde{\omega}_s}{\partial z}
                   +\beta\frac{G}{T}\right)
 +\frac{\phi_*D_{s*}}{k\,\phi\,D_s}
  \left(\frac{1}{\tilde{\omega}_{s*}}\frac{\partial\tilde{\omega}_{s*}}{\partial z}
        +\beta_*\frac{G_*}{T_*}\right)
 +\frac{u_{w0}}{k\chi\phi D_s}
     \left(1-\frac{\tilde{\omega}_{s*}}{\tilde{\omega}_s}\right)\,;
$$
and the parameter
$$
\beta\equiv\alpha_s(T)-\frac{\tilde{\mu}_sg}{RG}
$$
characterizes the strength of the non-Fickian flux.

With Eq.\,(\ref{eq210}), one can reconstruct the hydrate profile
$h(z)$ from the measured chlorinity profile $\tilde{\omega}_s(z)$.
Although one can write down an analytical solution to the problem
(\ref{eq210})
$$
h(z)=\int_0^zf(z_1)\,e^{-\int_{z_1}^z\gamma(z_2)\,dz_2}dz_1\,,
$$
numerical integration of Eq.\,(\ref{eq210}) is more convenient for
data analysis in practice. Remarkably, the relation between $h(z)$
and $\tilde{\omega}_s(z)$ does not involve quantitative data on
the process and history of the formation of hydrate deposit and
the process of generation of methane from sediments.

\begin{figure}[!t]
\center{
{\sf (a)}
\includegraphics[width=0.625\textwidth]%
 {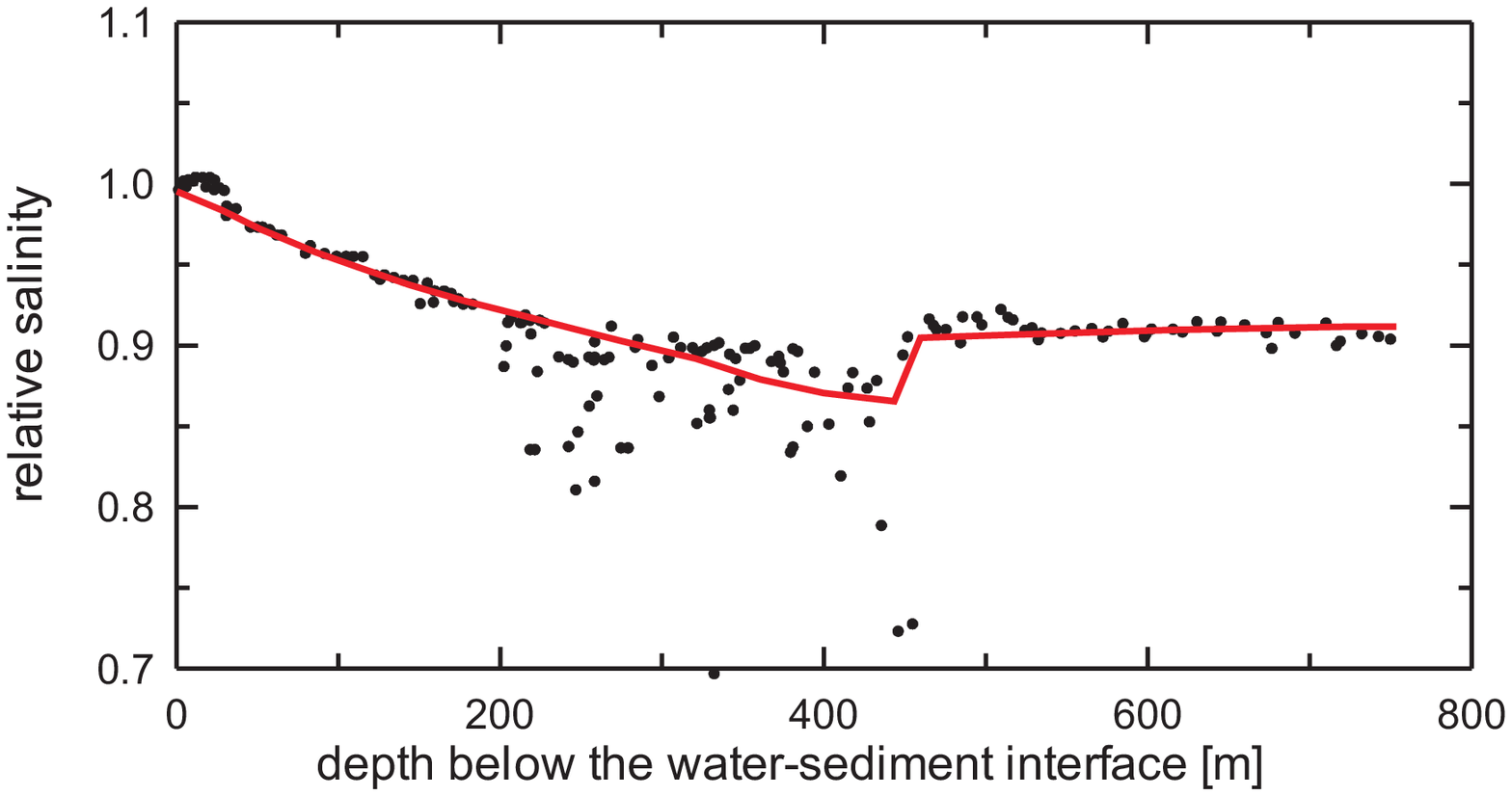}
\\[25pt]
{\sf (b)}
\includegraphics[width=0.625\textwidth]%
 {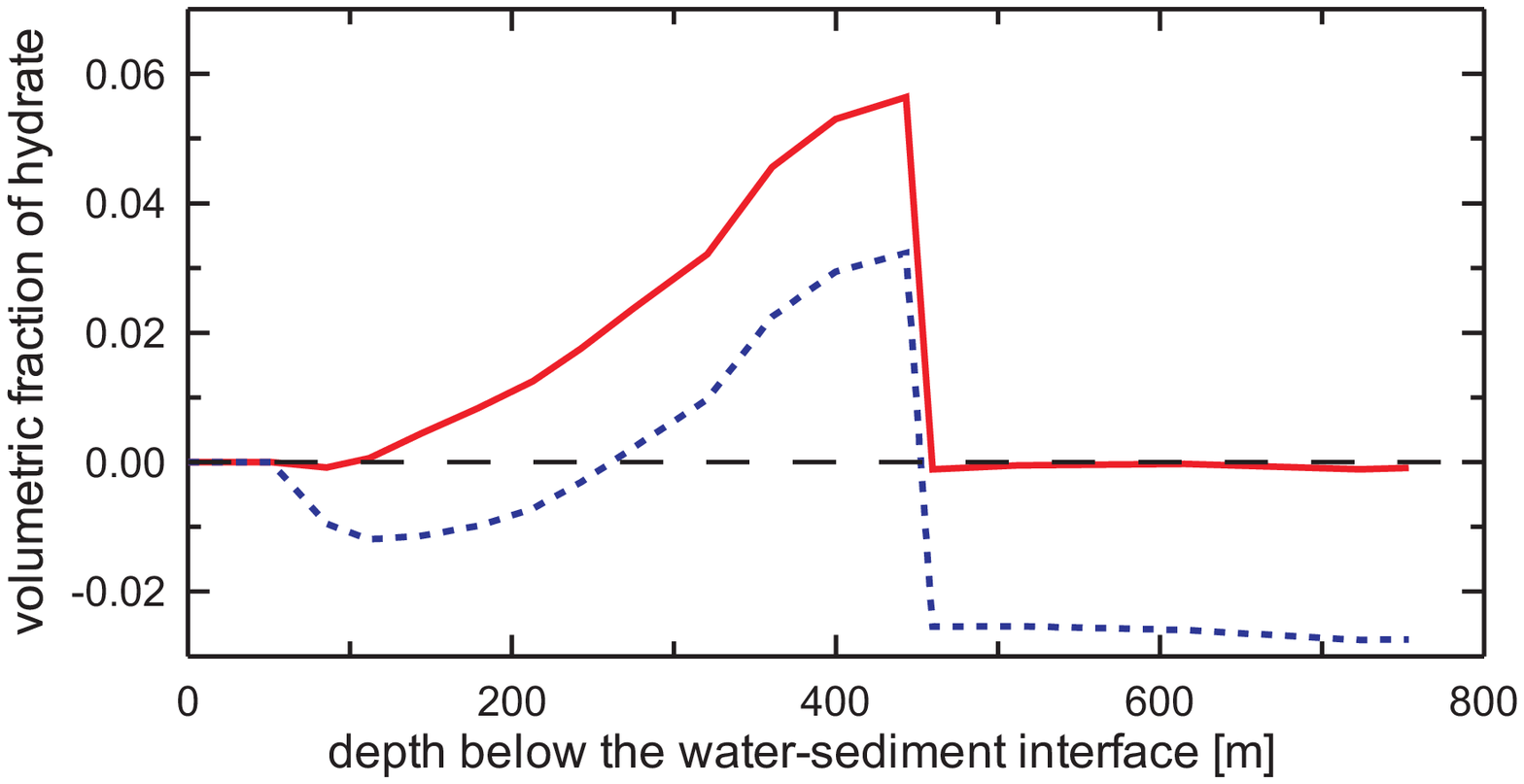}
}

  \caption{(Color online)
(a): Measured chlorinity profile (solid circles) and smoothed
chlorinity profile used for calculation of the hydrate profile
(red solid line) for the site 997 of the Ocean Drilling
Program~\cite{ODP}. (b): Hydrate profile reconstructed from the
chlorinity data with Eq.\,(\ref{eq210}) is plotted with the red
solid line (parameters are specified in Table~\ref{tab} and
$u_{w0}=-8\,\mathrm{cm/kyear}$, $v_{s0}=9\,\mathrm{cm/kyear}$).
For demonstration, we plot a {\it formal} hydrate profile for
purely Fickian diffusion flux of the same strength (blue dashed
line). Non-Fickian contributions are obviously non-negligible.
Indeed, the latter profile significantly deviates from the former
one and possesses unphysical features: negative values of the
hydrate volumetric fraction and non-zero (negative) amount of
hydrate beyond the zone of the thermodynamic stability of
hydrate.}
  \label{fig2}
\end{figure}

\section{SALINITY PROFILE ANALYSIS AND HYDRATE PROFILE RECONSTRUCTION}
We demonstrate application of our results to the analysis of one
of the most important marine hydrate provinces---the Blake Ridge.
For the Ocean Drilling Program site 997, on the Blake Ridge,
extensive data have been acquired, including hydrate samples and
the on-site salinity (chlorinity) measurements~\cite{ODP}
(Fig.\,\ref{fig2}(a)). The reported parameters for this site are
presented in Table~\ref{tab}. We have two parameters which are not
imposed by the results of direct measurements: sedimentation rate
$v_{s0}$ and filtration velocity $u_{w0}$.

\begin{table}[!t]
\caption{Geophysical properties for the Ocean Drilling Program site 997}
\begin{center}
\begin{tabular}{clcc}
\hline
\\[-5pt]
 $T_\mathrm{sf}$ & water-sediment interface temperature, Eq.\,(\ref{eq201}) &
   $2^\circ\mathrm{C}$ &  Refs.~\cite{ODP,Davie-Buffett-2001}\\[5pt]
 $G$ & geothermal gradient, Eq.\,(\ref{eq201}) &
   $35^\circ\mathrm{C/km}$ & Refs.~\cite{ODP,Davie-Buffett-2001}\\[5pt]
 $\phi(0)$ & seafloor porosity, Eq.\,(\ref{eq203}) &
   $0.69$ & Ref.~\cite{Davie-Buffett-2001}\\[5pt]
 $L$ & $e$-folding depth of porosity, Eq.\,(\ref{eq203}) &
   $2\,\mathrm{km}$ & Ref.~\cite{Davie-Buffett-2001}\\[5pt]
\hline
\end{tabular}
\end{center}
\label{tab}
\end{table}

In natural systems, the hydrate deposit cannot be in touch with
the water-sediment interface, because aqueous methane
concentration in the water body above the sediments is zero and
hydrate must dissociate. Moreover, in seas, methane is oxidized by
sulfates, which are present in sea water, and its concentration is
zero within the so-called sulfate reduction zone, which typically
expands approximately $20\,\mathrm{m}$ below the water-sediment
interface~\cite{Davie-Buffett-2003a}. Hence, hydrate should not be
present in a quiet extended upper part of the hydrate stability
zone. With Eq.\,(\ref{eq210}), the absence of hydrate, $h=0$,
requires $f(z)=0$ next to $z=0$. The function $f(z)$ is
independent of $v_{s0}$ and we can set it to zero for small $z$ by
tuning $u_{w0}$. With the chlorinity profile plotted in
Fig.\,\ref{fig2}(a), this procedure yields
$u_{w0}=-(8\pm0.5)\,\mathrm{cm/kyear}$ (the flux is negative,
i.e.\ ascending). One can see that for
$u_{w0}=-8\,\mathrm{cm/kyear}$, $h=0$ down to depths slightly over
$100\,\mathrm{m}$, while for different filtration velocity it
deviates from zero next to $z=0\,\mathrm{m}$. Furthermore, with
fixed $u_{w0}$, the reconstructed amount of hydrate below the
hydrate stability zone (for $z>450\,\mathrm{m}$ in
Fig.\,\ref{fig2}) depends on $v_{s0}$ monotonically; it vanishes
for $v_{s0}=(9\pm0.5)\,\mathrm{cm/kyear}$ (Fig.\,\ref{fig2}(b)).
The hydrate profile plotted in Fig.\,\ref{fig2}(b) with the red
solid line is the final result of the reconstruction of the
hydrate profile from the measured chlorinity profile plotted in
Fig.\,\ref{fig2}(a).

It is noteworthy that our reconstruction procedure is free of
uncertainties in parameters: all but two parameters are available
from direct measurements and these two parameters are strictly
imposed by two inevitable inherent features of the hydrate
profile.

\section{CONCLUSION AND DISCUSSION}
In this paper the transport of water and salt have been considered
for marine sediments hosting natural hydrate deposits. The
mathematical description employed accounts for\\
$\bullet$ non-Fickian diffusion of $\mathrm{NaCl}$, and\\
$\bullet$ temperature dependence of the molecular diffusivity.\\
We have demonstrated the crucial importance of the both, whereas
they are disregarded in the literature on the modelling of marine
hydrate deposits
(e.g.,~\cite{Davie-Buffett-2001,Davie-Buffett-2003a,Davie-Buffett-2003b,paradigm_models}).
Based on this consideration, we have derived the relation between
the measured salinity (chlorinity) profile and the hydrate
profile. Application of this relation has been demonstrated for a
real hydrate deposit (Fig.\,\ref{fig2}).

The solution, we found for this ``reverse engineering'' problem,
does not involve quantitative data on the process and history of
the formation of hydrate deposit and the process of generation of
methane from sediments. This is an important feature of our
results because previously, in the literature, closed models of
hydrate formation involve particular assumptions on the generation
process (e.g.,~\cite{Davie-Buffett-2001,Davie-Buffett-2003a}). In
these studies the entire model is tested against the measured
salinity profile, while we can see that only the current hydrate
profile determines the salinity profile. Moreover, the
sedimentation rate is unambiguously imposed by features of one of
these profiles, whereas it has been previously indirectly inferred
from geological data.

Importantly, our reconstruction procedure is free of uncertainties
in model parameters: all but two parameters---sedimentation rate
$v_{s0}$ and filtration velocity $u_{w0}$---are available from
direct measurements. These two parameters are strictly imposed by
two inherent features of the hydrate profile: the absence of
hydrate (i) close to the water-sediment interface and (ii) beneath
the hydrate stability zone.

\section*{ACKNOWLEDGEMENTS}
The work has been financially supported by the Government of Perm
Region (Contract C-26/212) and Grant of The President of Russian
Federation (MK-6932.2012.1).

\bibliographystyle{elsarticle-num}

\end{document}